\documentclass[twocolumn,twoside,slac_two]{revtex4}
\usepackage{graphicx}
\usepackage{fancyhdr}
\newcommand{\dzero}     {D0}

\newcommand{\ttbar}     {\mbox{$t\bar{t}$}}

\begin{document}

\title{Top quark properties measurement with the $D0$ detector}

%

\author{Shabnam Jabeen on behalf of D0 collaboration}
\affiliation{Department of Physics, Boston University, Boston MA, USA.}

\begin{abstract}
 One of the main goals of the Tevatron RunII is to look for any hints for
 new physics. At D0, the range of searches for new physics signals is
large and one of the places we look for hints for new physics is by
measuring the top quark properties. A few of these measurements are
 discussed in this paper.

\end{abstract}

\maketitle

\thispagestyle{fancy}


\section{Introduction}
Tevatron, a $p\bar p$ collider with center of mass energy of about 2 TeV, is
 currently world's highest energy collider in operation.
It has been running quite efficiently for the last many years delivering more
 than $6~fb^{-1}$ integrated luminosity in RunII so far. The D0 detector
 has also been
performing excellently. The motivation to accumulate as much
 luminosity as possible is
 not only to find Higgs and do precision measurements but also to look for new
physics signals and rule out as many physics models as one can.
 The top quark is by far the heaviest fermion in the standard model, and
 thus has the strongest coupling to the Higgs boson of all standard model 
fermions.
This makes the top quark and its interactions an ideal place to look for
 new physics related to electroweak symmetry breaking.
In this paper, we present three analyzes which look for new physics
 signals in the top quark sector.

\section{Measurement of CKM matrix element $V_{tb}$}

Within the standard model the top quark decays
to a W boson and a down-type quark q (q = d; s; b) with
a rate proportional to the squared Cabibbo-Kobayashi-
Maskawa (CKM) matrix element~\cite{CKM}. Under the
assumption of three fermion families and a unitary $3x3$
CKM matrix, the $V_{tq}$ elements are severely constrained~\cite{CKM2}.
 However, in the presence of new physics CKM submatrix may not
be a 3X3 unitary matrix and in that case $V_{tq}$ elements
 can significantly deviate from their standard model values.
 This would affect, among other things, the ratio R of the top quark branching
fractions, which can be expressed in terms of the CKM
matrix elements as

\begin{eqnarray*}
\label{eq:Rdef}
R = \frac{{ \cal B}(t \rightarrow Wb)}{{ \cal B}(t \rightarrow Wq)} & = &
\frac{\mid V_{tb}\mid^2}{\mid V_{tb}\mid^2 + \mid V_{ts}\mid^2 +
 \mid V_{td}\mid^2}  \;.
\end{eqnarray*}
Thus by measuring R precisely we can set limit on the ratio of $|V_{tb}|^2$
to the off-diagonal  matrix elements without any assumptions on the
unitarity of the CKM matrix.
The analysis presented here is based on data collected with the D0 detector
\cite{run2det} between August 2002 and December 2005 at the Fermilab Tevatron
{\mbox{$p\bar p$}}\ collider at {\mbox{$\sqrt{s}$ =\ 1.96\ TeV}},
corresponding to an integrated luminosity of
about $0.9~\text{fb}^{-1}$. The analysis uses the top quark pair production.
 Within standard model top quarks decay to a $W$ boson and a $b$ quark 
almost 100\% of the time. For this analysis we only consider events in 
which one of the $W$ bosons decays  into two quarks, and the other  
one into an electron or muon and a neutrino.

\begin{figure}[!h!btp]
\includegraphics[width=0.48\textwidth]{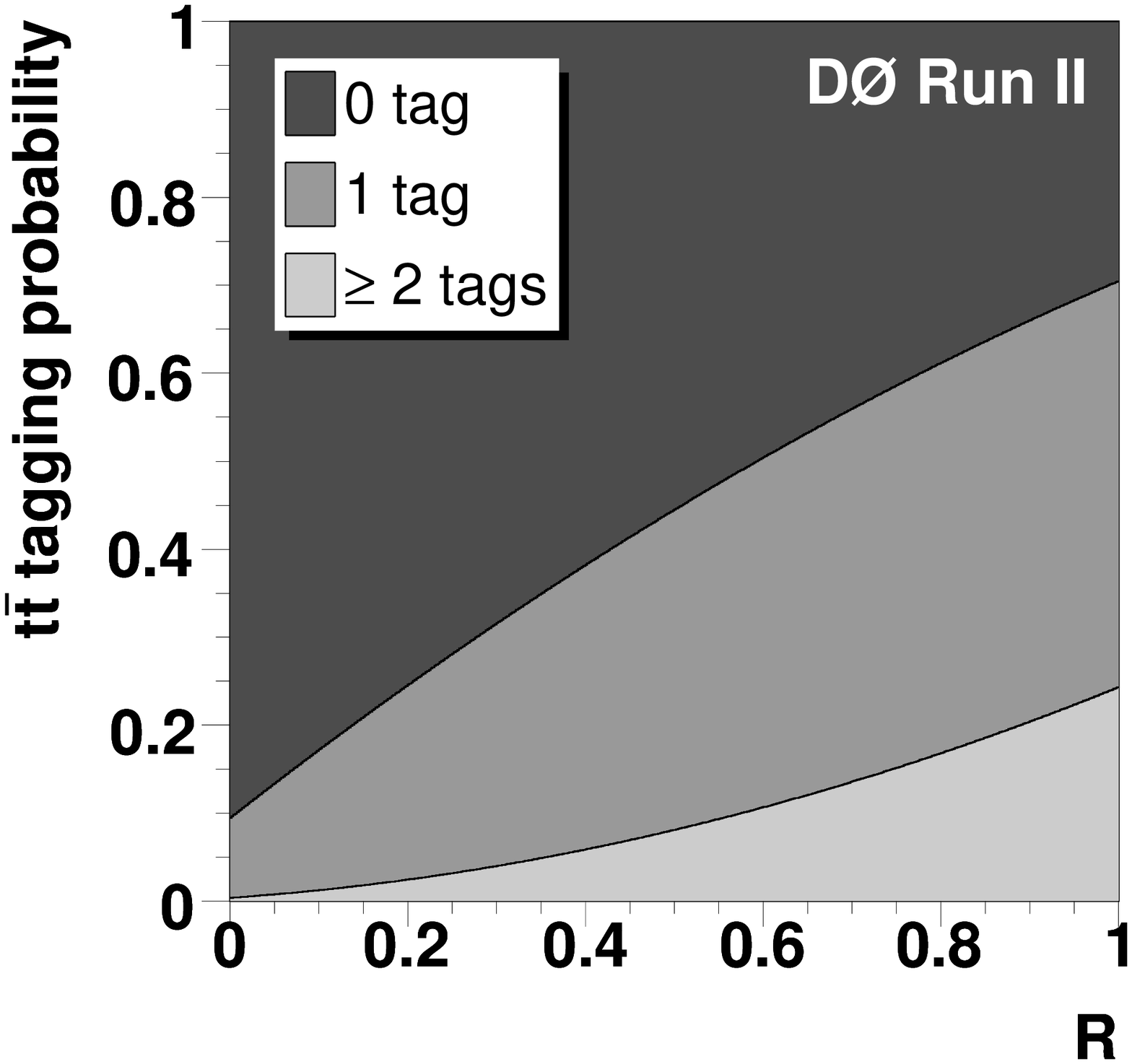}
\includegraphics[width=0.48\textwidth]{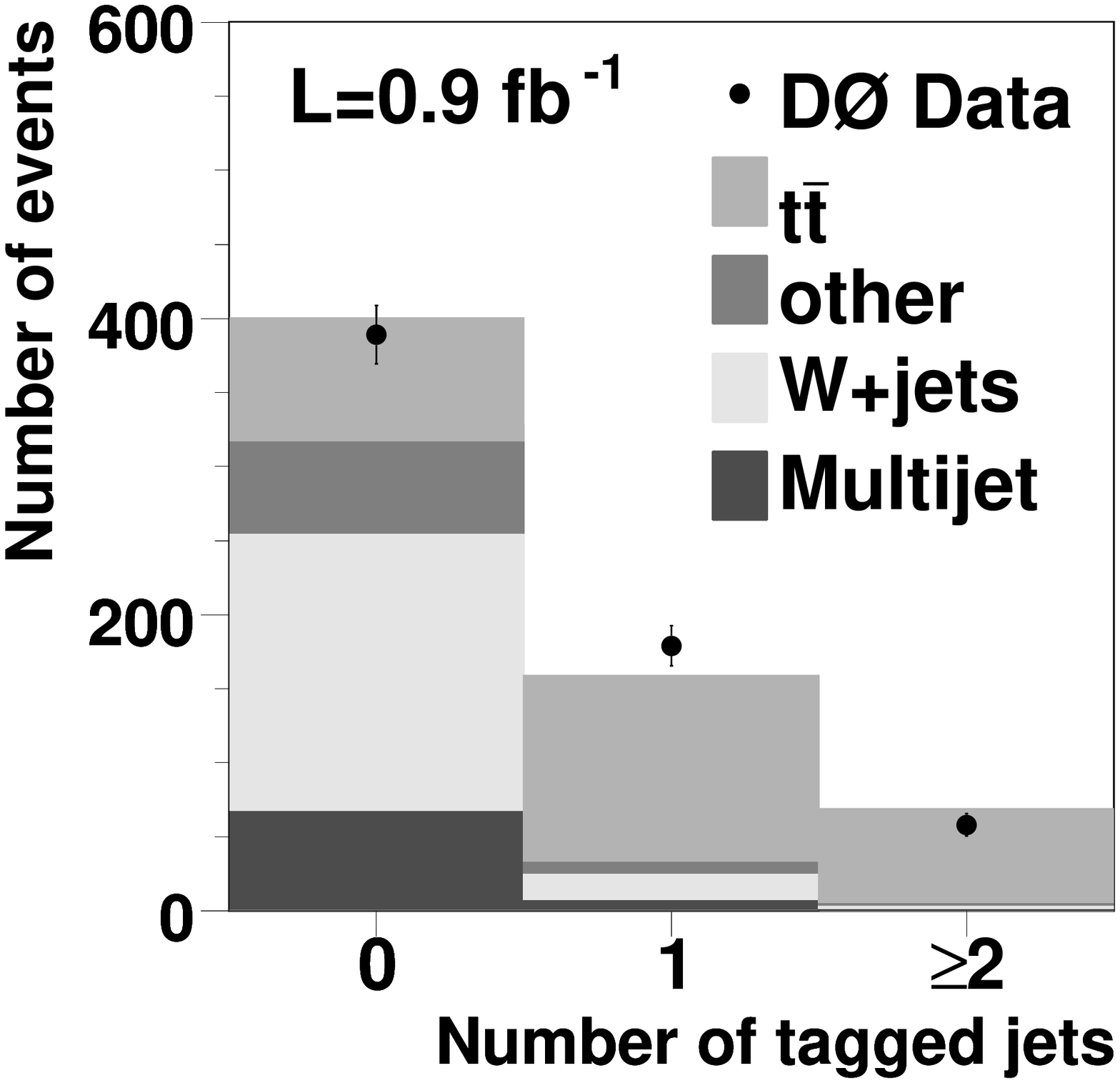}
\caption{Above: Fractions of events with 0, 1 and $\ge 2$
b tags as a function of R for  $t \bar t$ events with $\ge 4$ jets;
Below: predicted and observed number of events in the 0,
1 and $\ge 2$ b tag samples for the measured R and $t \bar t$ for
events with $\ge 4$ jets.
}
\label{fig:R1}
\end{figure}

We identify $b$-jets using a neural-network tagging algorithm.
We split the selected sample into subsamples according
to the lepton flavor ($e$ or $\mu$), jet multiplicity (3 or $\ge 4$ jets) and
number of identified $b$-jets (0, 1 or $\ge 2$), thus obtaining 12 disjoint
data sets.  
Since the probability to tag a \ttbar~ event depends on the flavor of the jets,
it depends on $R$.
We estimate the acceptance and tagging probabilities
for each of the three {\ttbar} decay modes $bb$, $bq_l$ and $q_lq_l$.
Figure~\ref{fig:R1} shows tagging probabilities as a function of $R$
for \ttbar~ events with $\ge 4$ jets and 0, 1 and
$\ge 2$ $b$ tags.

\begin{figure}[!h!btp]
\includegraphics[width=0.48\textwidth] {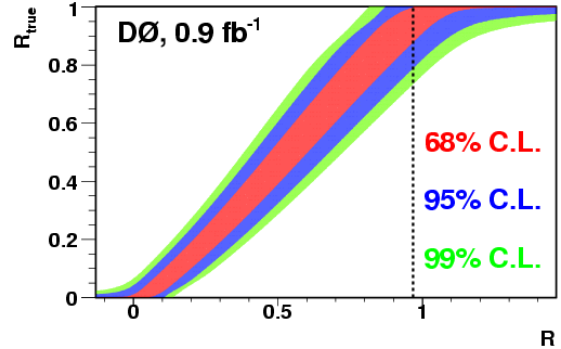}
\caption{The 68\%, 95\%, and 99\%
C.L. band for $R_{true}$ as a function of
$R$. The dotted black line indicates the measured value
$R = 0.97$.
}
\label{fig:R2}
\end{figure}

\section{Anomalous $Wtb$ couplings}

Anomalous $Wtb$ couplings modify the angular correlations of the
 top quark decay
products and change the single top quark production cross section.
 In this paper we also present the first study of $Wtb$ couplings
that combines $W$~helicity measurements in top quark decay with anomalous
 coupling searches in the single top quark final state.

The effective Lagrangian describing the $Wtb$
interaction including operators up to dimension five is:
\begin{eqnarray}
\mathcal{L}&=&-\frac{g}{\sqrt{2}}\bar{b} \gamma^\mu V_{tb}
\nonumber (f^L_1 P_L + f^R_1 P_R) t W_{\mu}^{-}\\
&-&\frac{g}{\sqrt{2}} \bar{b} \frac{i\sigma^{\mu\nu} q_{\nu}V_{tb}}{M_W}
 (f^L_2 P_L + f^R_2 P_R) t W_{\mu}^{-}
+ h.c. \, ,
\label{coupling}
\end{eqnarray}

where  $M_W$ is the mass of the $W$~boson, $q_{\nu}$ is its four-momentum,
 $V_{tb}$ is the
Cabibbo-Kobayashi-Maskawa matrix element, and $P_{L}=(1 - \gamma_5)/2$
($P_{R}=(1 + \gamma_5)/2$) is the left-handed (right-handed) projection 
operator.  In the standard model, the $Wtb$ coupling is purely left-handed,
and the values
of the coupling form factors are $f^{L}_{1} \approx 1 $,
 ${f^{L}_{2}=f^{R}_{1}=f^{R}_{2}=0}$. For this analysis we assume real 
coupling form factors, implying $CP$ conservation, and a spin-$\frac{1}{2}$
top quark which decays predominantly to $Wb$.

We investigate one pair of coupling form factors at a time and consider
three cases, pairing the left-handed vector coupling form factor
$f^{L}_{1}$ with each of the other three form factors. We refer to
these as $(L_1,R_1)$, $(L_1,L_2)$, and $(L_1,R_2)$. For each
 pair under investigation we assume that the other two have 
the standard model values.

In this analysis we combine information
from our measurement of the $W$~boson helicity fractions in $t\bar{t}$
events~\cite{Whel-d0} with information from single top quark production.
We have set direct limits on anomalous top quark coupling before
~\cite{Wtb-prl} but those limits are based on single top
quark final states only. This new measurement
is based on a sample of 0.9 fb~$^{-1}$ of single top
candidate events and up to 2.7 fb~$^{-1}$ of $t\bar{t}$ candidates
collected by the {\dzero} detector.

\begin{figure}[!h!btp]
\includegraphics[width=0.48\textwidth]{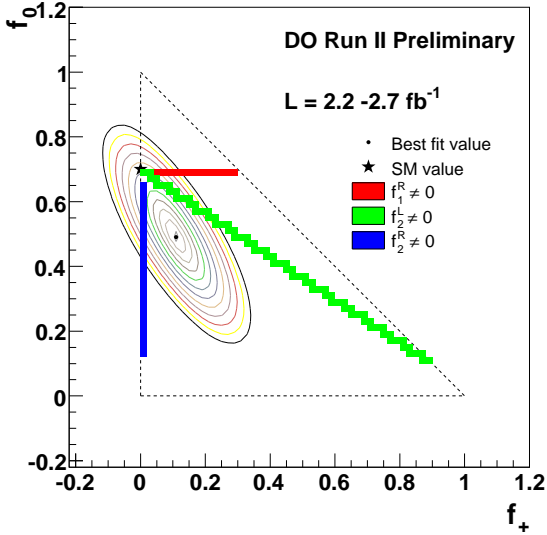}
\vspace{-0.1in}
\caption[feynman]{Graphical representation of the change in $W$
boson helicity fractions away from the SM values (shown
by the star) if the anomalous couplings are present.}
\label{fig:AC1}
\end{figure}

The W boson helicity measurement, described in
Ref.~\cite{Whel-d0}, uses events in both the $l$+jets 
$(t \bar t \rightarrow W^+W^- b \bar b \rightarrow l \nu q 
\bar q b \bar b )$ and dilepton $(t \bar t \rightarrow W^+W^- 
b \bar b \rightarrow l \nu l' \nu '  b \bar b )$ final states, 
and is extracted form the distribution
of $\theta ^*$, the angle between the down-type
fermion and top quark momenta in the W boson rest
frame. For each pair of form factors a likelihood distribution
is extracted from the W helicity measurement
of the decay angle distribution in top quark decays.
We vary both the longitudinal and right-handed helicity fractions f$_0$
and f$_+$ in the fit and find the relative likelihood of any
set of helicity fractions being consistent with the data.
The result is presented in Fig.~\ref{fig:AC1}, which also demonstrates
how non-SM values for the coupling form factors
alter the W helicity fractions.

\begin{figure}[!h!btp]
\includegraphics[width=0.4\textwidth]{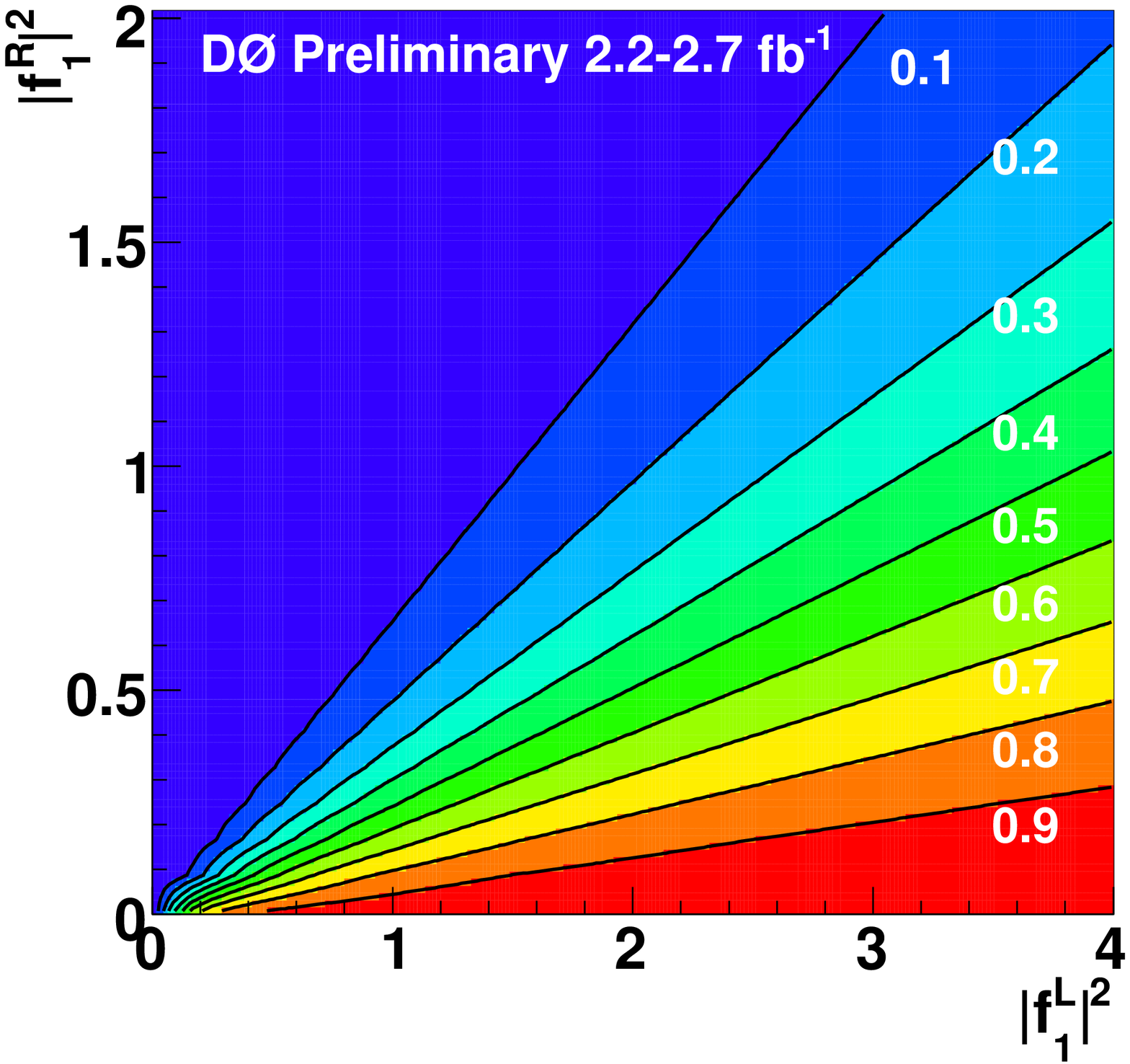}
\includegraphics[width=0.4\textwidth]{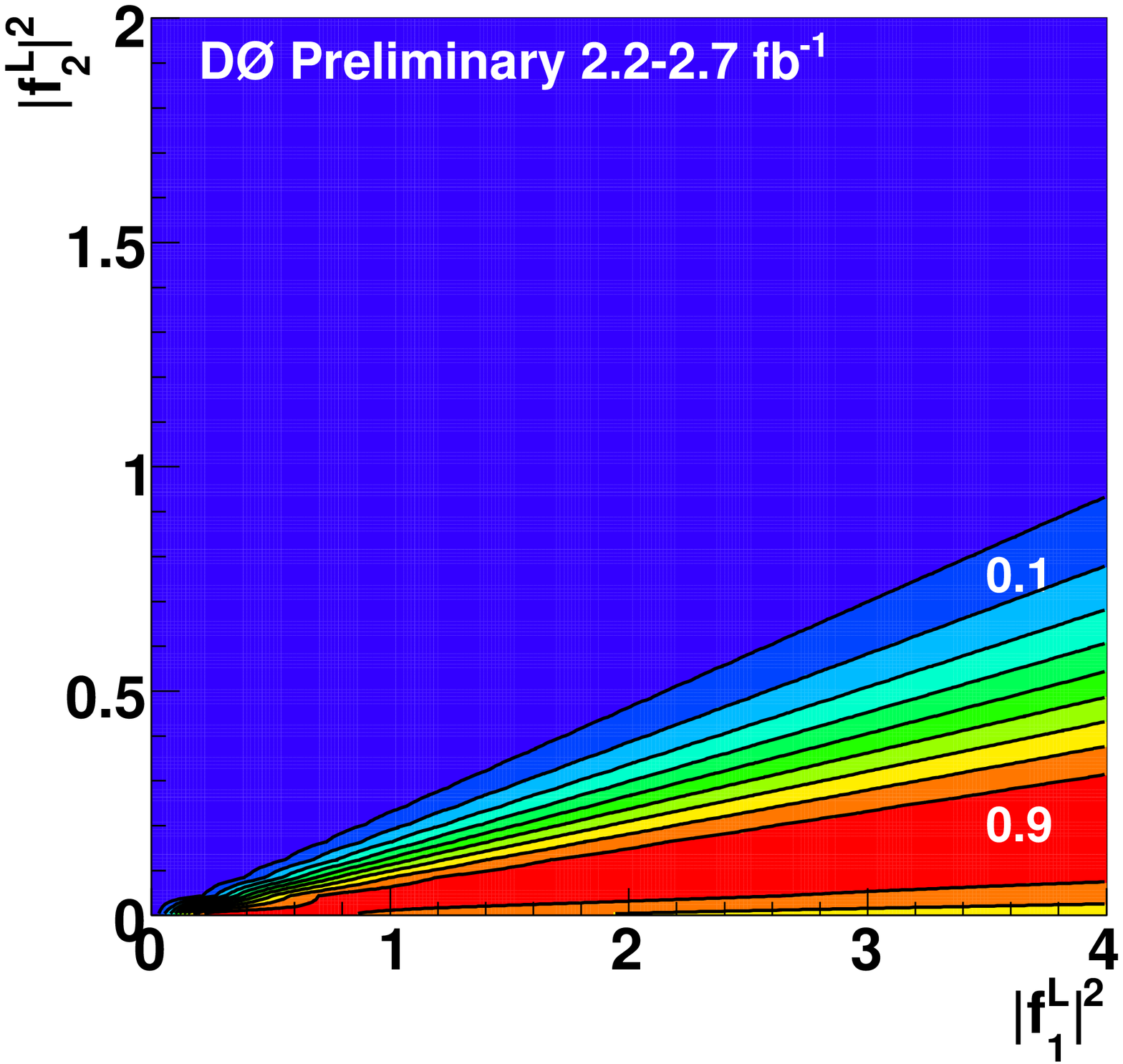}
\includegraphics[width=0.4\textwidth]{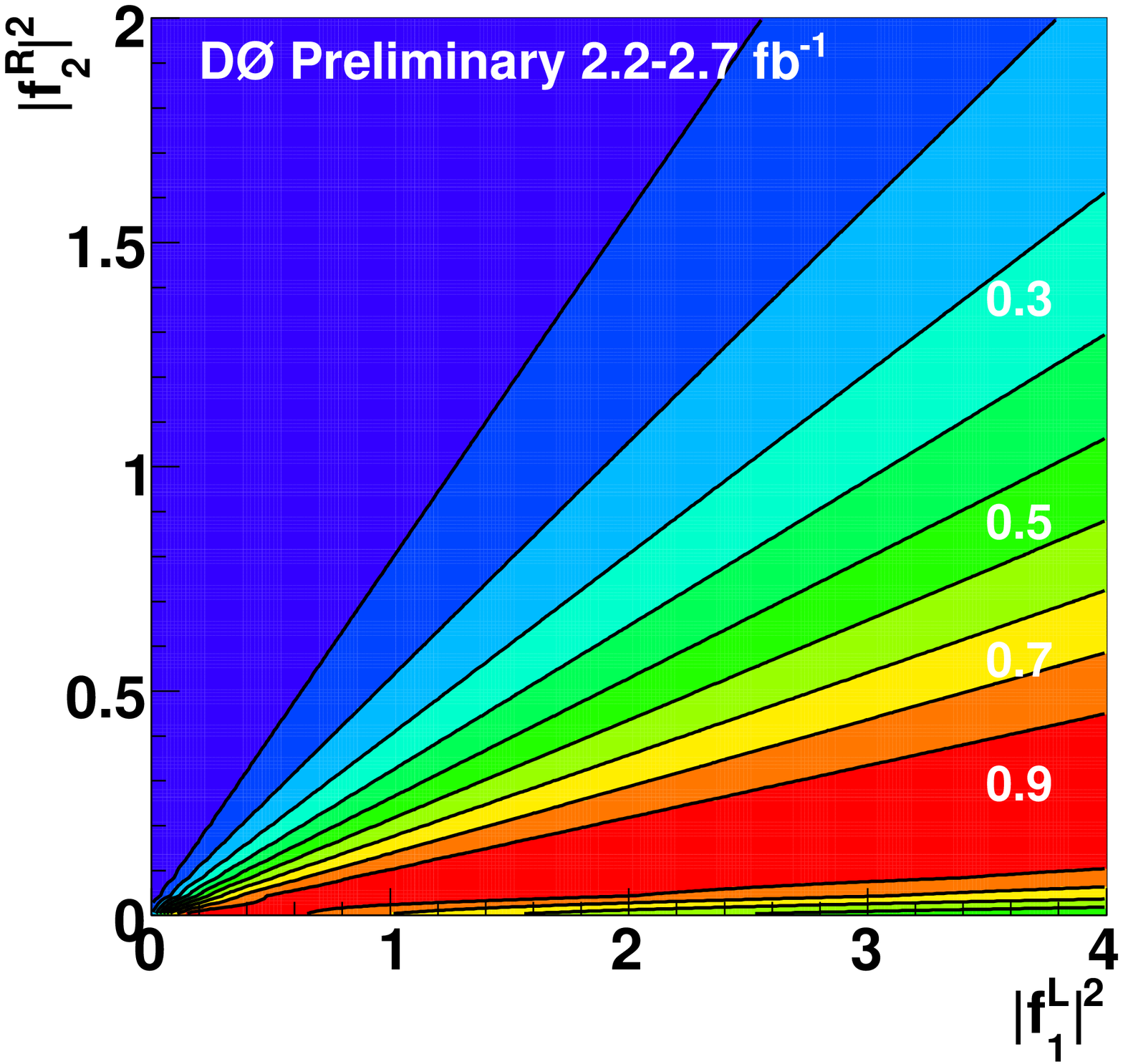}
\caption{$W$~helicity prior  for right- vs left-handed vector coupling (a),
left-handed tensor vs left-handed vector coupling (c), and right-handed tensor
 vs left-handed vector coupling (e). The $W$~helicity prior is normalized
to a peak value of one and shown as equally spaced contours
 between zero and one.
}
\label{fig:measfullsys_2D_prior}
\end{figure}

This likelihood from the W helicity analysis is
then used as a prior in a Bayesian statistical analysis for the anomalous 
coupling search in single top
quark production and decay channels, yielding a two dimensional
posterior probability density as a function
of both form factors. We extract limits on $f_1^R$, $f_2^L$, and $f_2^R$
 by projecting the two-dimensional posterior
onto the corresponding form factor axis. The
W boson helicity measurement is described in Ref.~\cite{Whel-d0}
and the helecity priors are shown in Fig.~\ref{fig:measfullsys_2D_prior}.

\begin{figure}[!h!btp]
\includegraphics[width=0.48\textwidth]{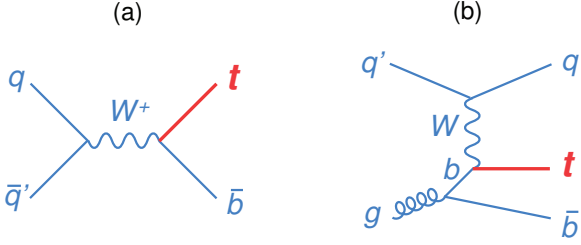}
\vspace{-0.1in}
\caption[feynman]{Feynman diagrams for (a) $s$-channel
and (b) $t$-channel single top quark production.}
\label{feynman-diagrams}
\end{figure}

\begin{figure}[!h!tbp]
\includegraphics[width=0.48\textwidth]{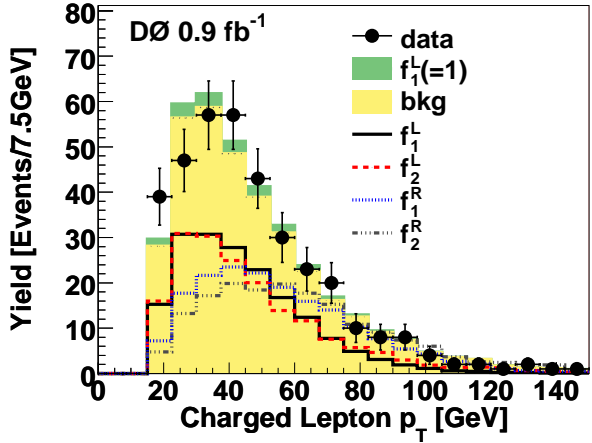}
\vspace{-0.1in}
\caption[lep-pt]{Charged lepton $p_T$ spectrum from data and
expectation for SM single top production plus background for
events with two jets, one $b$-tagged jet. Superimposed are the
 distributions from single top quark production with different couplings
 (all other couplings set to zero) normalized to ten times the SM single
 top quark cross section.}
\label{lep-pt_sum}
\vspace{-.1in}
\end{figure}

The dominant tree level Feynman diagrams for single top quark production 
in $p \bar p$ collisions are illustrated in Fig.~\ref{feynman-diagrams}.
 In this analysis we combine both these production modes and assume 
that single top quark production proceeds exclusively through $W$ 
boson exchange.  The presence of anomalous couplings can change 
angular distributions and event kinematics as demonstrated by the
 $p_T$ spectrum of the charged lepton from the decay of the top 
quark in Fig.~\ref{lep-pt_sum}. Such differences can be used to 
distinguish these couplings. We use boosted decision trees to 
discriminate between the single top quark signal and background.

We use Bayesian statistics to compare the output distribution of the decision 
trees from data to expectations for single top quark production. For any pair
 of values of the two couplings that are considered non-zero, we compute the 
expected output distribution by superimposing the distributions from the two
 signal samples  with the non-standard coupling and from the background 
samples in the appropriate proportions. In case of the $(L_1,L_2)$ scenario, 
the two amplitudes interfere, and we use a superposition of three signal
 samples, one with left-handed vector couplings, one with the left-handed 
tensor coupling only set to one, and one with both couplings set to one to
 take into account the effect of the interference. We then compute a 
likelihood as a product over all bins and channels. Here we use twelve 
channels defined by lepton flavor, $b$~tag multiplicity, and jet multiplicity 
(2, 3, or 4).

\begin{figure}[!h!btp]
\includegraphics[width=0.4\textwidth] {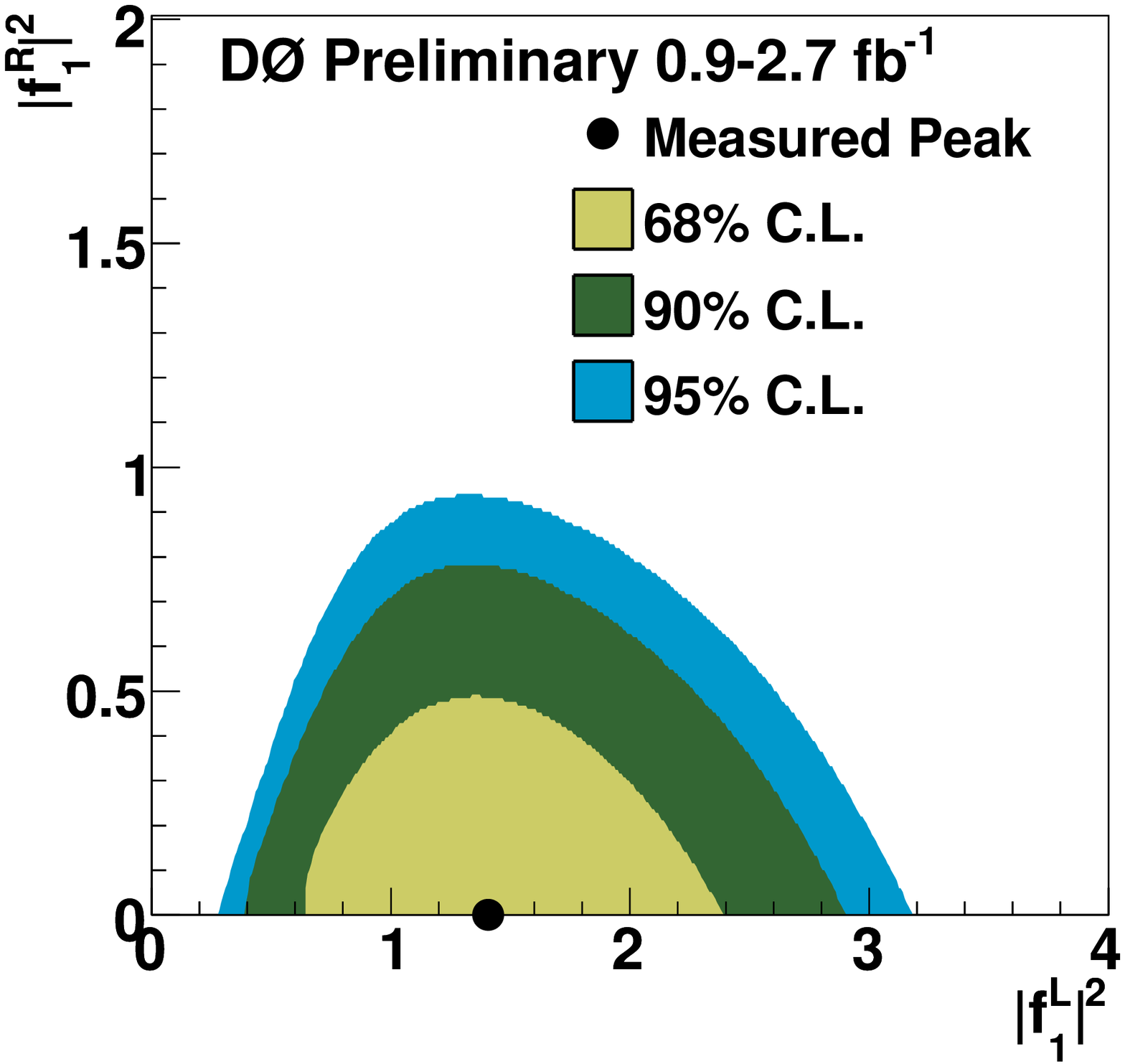}
\includegraphics[width=0.4\textwidth] {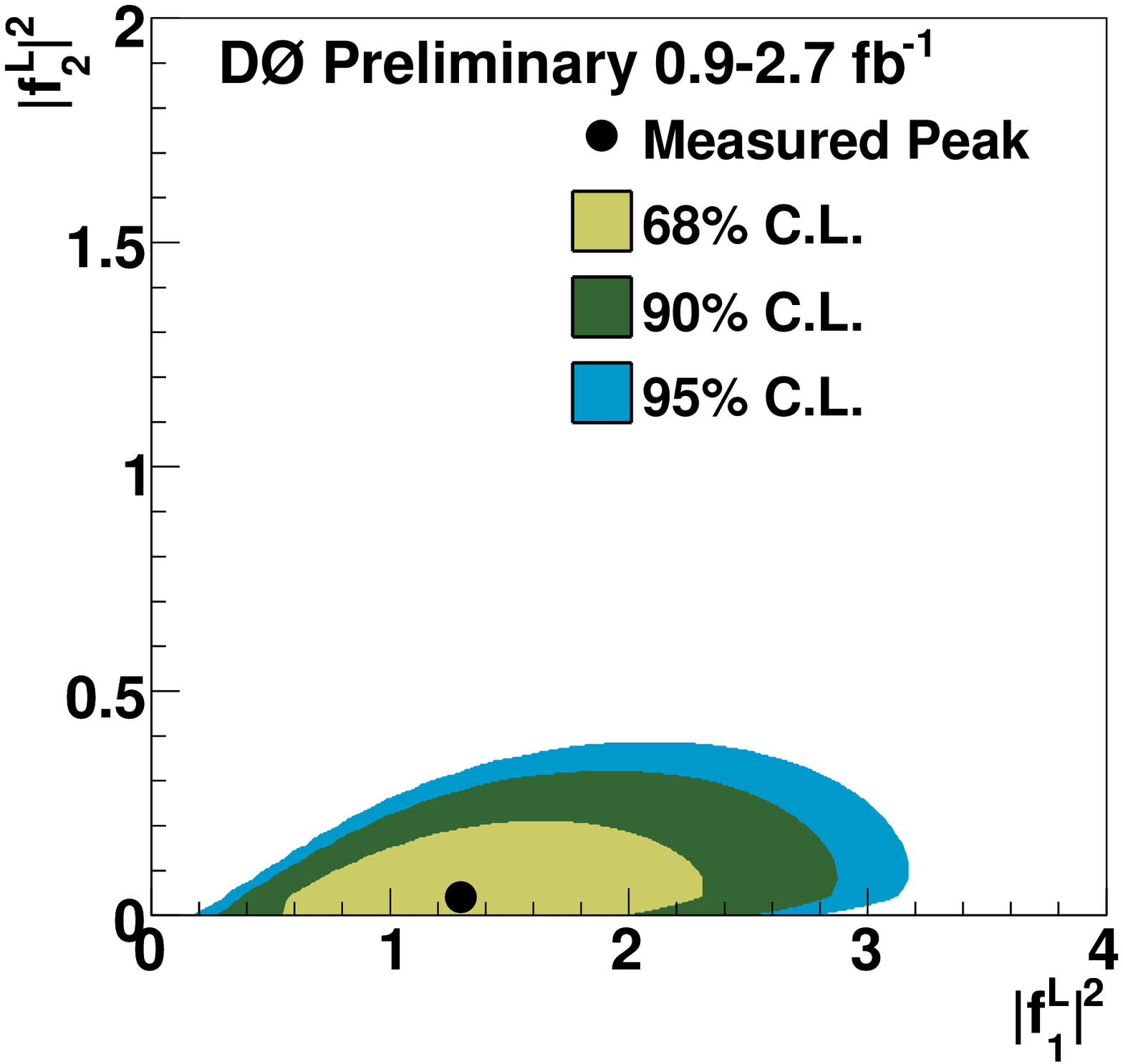}
\includegraphics[width=0.4\textwidth] {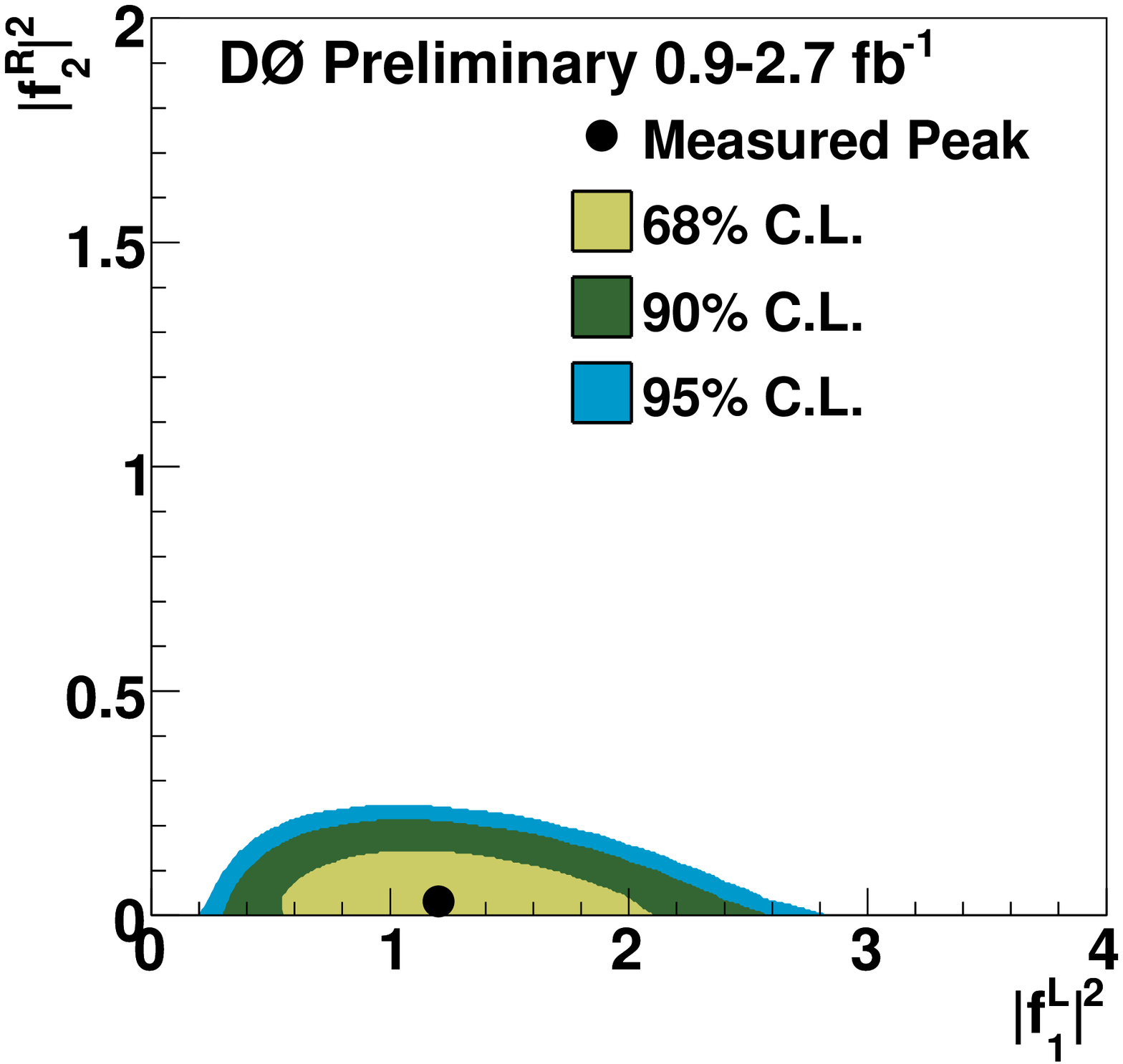}
\caption{final posterior density for right- vs left-handed vector coupling (b),
 left-handed tensor vs left-handed vector coupling (d), and right-handed tensor
 vs left-handed vector coupling (f).  The posterior density is shown as
 contours of equal probability density.
}
\label{fig:measfullsys_2D}
\end{figure}

The two-dimensional posterior probability density is computed as a function of $|f^{L}_{1}|^2$ and
$|f_X|^2$, where $f_X$ is $f^{R}_{1}$, $f^{L}_{2}$, or $f^{R}_{2}$.
These probability distributions are shown in~Fig.~\ref{fig:measfullsys_2D}. In all three
scenarios we measure approximately zero for the anomalous coupling form factors and favor the
left-handed vector hypothesis over the alternative hypothesis.
We compute 95\%~Confidence Level (C.L.) upper limits on these form factors by integrating out the left-handed vector
coupling form factor to get a one-dimensional posterior probability density. The measured values are
given in Table~\ref{table:obslim}.

\begin{table}[t]

\caption{\label{table:obslim} Measured values with uncertainties and upper 
limits at the 95\%~C.L.
for $Wtb$ couplings in three different scenarios.}
\begin{tabular}{l@{ }l@{ }l} \hline\hline
Scenario  & ~Coupling                   & Coupling limit if $f^L_1=1$    \\
\hline    
\vspace{-0.08in} \\
($L_1,R_1$)    &  $|f^L_1|^2=1.27 ^{+0.57}_{-0.48}$      &   \\
               &  $|f^R_1|^2<0.95 $                      & $|f^R_1|^2< 1.01 $ \\
($L_1,L_2$)    &  $|f^L_1|^2=1.27 ^{+0.60}_{-0.48}$      &   \\
               &  $|f^L_2|^2<0.32 $                      & $|f^L_2|^2< 0.28 $ \\
($L_1,R_2$)    &  $|f^L_1|^2=1.04 ^{+0.55}_{-0.49}$      &   \\
               &  $|f^R_2|^2<0.23 $                      & $|f^R_2|^2< 0.23 $ \\
\hline \hline
\end{tabular}
\vspace{-.1in}
\end{table}

\section{Top- antitop spin correlations}

\begin{figure}[!h!btp]
\includegraphics[width=0.48\textwidth] {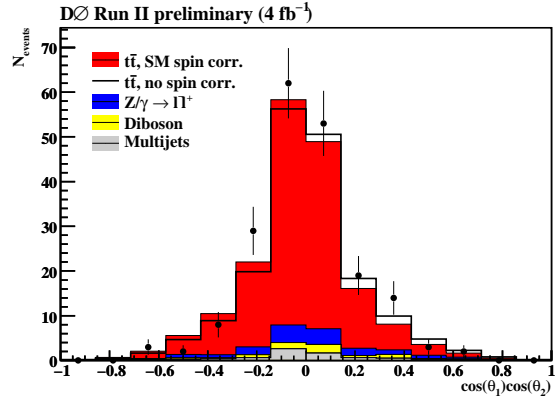}
\caption{The $cos(\theta _1)cos(\theta_2)$ distribution for
full dilepton event sample. The sum of $t \bar t$ signal including NLO QCD
 spin correlation ($C = 0.777$) and multijet, diboson
 and Drell-Yan  background is compared to data.
The open black histogram shows the prediction without
 $t \bar t$ spin correlation ($C=0$);
}
\label{fig:SpinCorr1}
\end{figure}

\begin{figure}[!h!btp]
\includegraphics[width=0.48\textwidth] {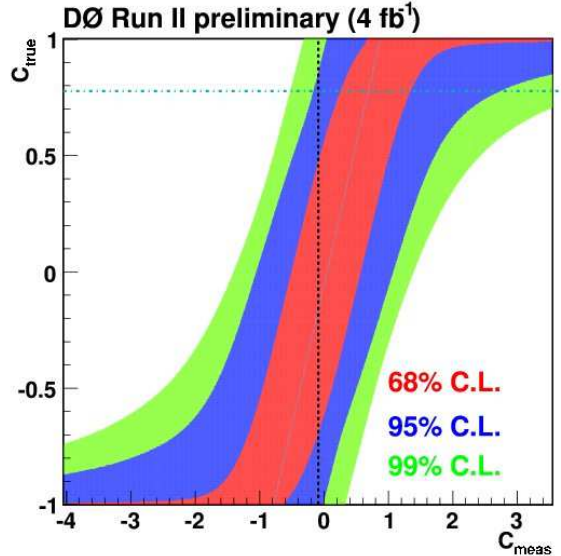}
\caption{The 68\%, 95\%, and 99\%
C.L. band for $C_{true}$ as a function of $C_{meas}$. The
thin, slanted line indicates the most probable value of $C_{true}$ as
 a function of $C_{meas}$ and represents therefore the bias of the method.
The vertical black line depicts the measured value $C_{meas} = -0.090$.
The horizontal line indicates the NLO QCD
value $C_{true} = 0.777$.
}
\label{fig:SpinCorr2}
\end{figure}

Top quark physics plays an important role in testing the atandard model
and its possible extensions. One of the most important properties
 of the top quark, the spin, has not been carefully explored.
While the top quarks and antiquarks produced at hadron colliders
 are unpolarized, their spins are correlated.
Since at the Tevatron top pair production is dominated by
 $q \bar q$ scattering, a different spin correlation is analyzed
compared to the LHC where top pair production is dominated by
 $gg$ scattering. The standard model predicts that the top quark
decays before its spin flips, in contrast with the lighter
quarks, which are depolarized by QCD interactions long
before they fragment. The spin of the top quark is
 therefore reflected by its decay products. In this analysis
it is assumed that top quarks decay exactly as predicted by the standard model.
Then the charged lepton from a leptonic top
quark decay has a spin analyzing power of 1 at the tree level.
Therefore, the dilepton final states have the highest
sensitivity to measure the correlation between the spins of
 pair-produced top and anti-top quarks.
The observation of spin correlations would result in an
upper limit on the lifetime of the top quark. This can be
translated into a lower limit on the Kobayashi-Maskawa matrix
element $|V_{tb}|^2$ without making assumptions about the
number of quark generations. Moreover, many scenarios beyond
the standard model predict different production and
decay dynamics of the top quark, which could affect the observed spin
correlation.
In the analysis presented here, the double differential angular
 distribution is used. The double differential distribution
 for a measurement of spin correlations between top and  and antitop quark 
can be expressed as:
\begin{eqnarray*}
\frac{1}{\sigma} \frac{d \sigma}{d cos \theta_1 cos \theta_2}= \frac {1}{4}
 (1 - C cos \theta_1 cos \theta_2 ),
\end{eqnarray*}
where $\sigma$ denotes the cross section of the channel under consideration
 and $C$ is a free parameter between -1 and 1 that
depends on the choice of the spin basis. For this analysis the beam axis was
 chosen to be the spin quantization axis for which the value for the
 coefficient constant including NLO QCD corrections is $C=0.777$.
For this measurement, we analyze the dileptonic
decay channels where the W bosons from the top and antitop
 quark decay into an electron and an electron neutrino
or into a muon and a muon neutrino. The $ee$ channel with 1.1
fb~$^{-1}$, the $e \mu$ channel with 4.2 fb~$^{-1}$ and the $\mu \mu$ channel
with 1.1 fb~$^{-1}$ of integrated luminosity are analyzed separately
and are then combined. The $cos \theta_1 cos \theta_2 $ distribution for
 the full sample is shown in Fig.~\ref{fig:SpinCorr1}. A likelihood fit 
for data gives a measured value for the spin correlation parameter
$C_{meas} = -0.09^{+0.59}_{-0.58}$ (stat + syst).
The calibrated spin correlation coefficient $C_{meas} = -0.17^{+0.64}_{-0.53}$
(stat + syst) has been measured using Feldman-Cousins procedure and is shown
  shown in Fig.~\ref{fig:SpinCorr2}.
This agrees with the standard model prediction for a spin 1/2 top
 quark of $C = 0.777$ in NLO QCD within 2 standard deviations.

For analysis details please see Ref.~\cite{SC}.

\begin{acknowledgments}
%
We thank the staffs at Fermilab and collaborating institutions, 
and acknowledge support from the 
DOE and NSF (USA);
CEA and CNRS/IN2P3 (France);
FASI, Rosatom and RFBR (Russia);
CNPq, FAPERJ, FAPESP and FUNDUNESP (Brazil);
DAE and DST (India);
Colciencias (Colombia);
CONACyT (Mexico);
KRF and KOSEF (Korea);
CONICET and UBACyT (Argentina);
FOM (The Netherlands);
STFC and the Royal Society (United Kingdom);
MSMT and GACR (Czech Republic);
CRC Program, CFI, NSERC and WestGrid Project (Canada);
BMBF and DFG (Germany);
SFI (Ireland);
The Swedish Research Council (Sweden);
and
CAS and CNSF (China).
%
\end{acknowledgments}

\bigskip 

\end{document}